# Optimal Scheduling of Energy Storage for Power System with Capability of Sensing Short-Term Future PV Power Production

Sarvar Hussain Nengroo
*The Cho Chun Shik Graduate School of Green Transportation*
*Korea Advanced Institute of Science and Technology*
Daejeon 34141, South Korea
sarvar@kaist.ac.kr

Sangkeum Lee
*Environment ICT Research Section*
*Electronics and Telecommunications Research Institute (ETRI)*
Daejeon 34141, South Korea
sangkeum@etri.re.kr

Hojun Jin
*The Cho Chun Shik Graduate School of Green Transportation*
*Korea Advanced Institute of Science and Technology*
Daejeon 34141, South Korea
hjjin1995@kaist.ac.kr

Dongsoo Har
*The Cho Chun Shik Graduate School of Green Transportation*
*Korea Advanced Institute of Science and Technology*
Daejeon 34141, South Korea
dshar@kaist.ac.kr

*Abstract*—Constant rise in energy consumption that comes with the population growth and introduction of new technologies has posed critical issues such as efficient energy management on the consumer side. That has elevated the importance of the use of renewable energy sources, particularly photovoltaic (PV) system and wind turbine. This work models and discusses design options based on the hybrid power system of grid and battery storage. The effects of installed capacity on renewable penetration (RP) and cost of electricity (COE) are investigated for each modality. For successful operation of hybrid power system and electricity trading in power market, accurate predictions of PV power production and load demand are taken into account. A machine learning (ML) model is introduced for scheduling, and predicting variations of the PV power production and load demand. Fitness of the ML model shows, when employing a linear regression model, the mean squared error (MSE) of 0.000012, root mean square error (RMSE) of 0.003560 and $R^2$ of 0.999379. Using predicted PV power production and load demand, reduction of electricity cost is 37.5 % when PV and utility grid are utilized, and is 43.06% when PV, utility grid, and storage system are utilized.

## I. INTRODUCTION

The energy demand in developing countries is increasing every day, and global energy demand will increase by 30% of its current level by 2050, based on an expected population increase [1]. Many countries have performed extensive research on renewable energy resources in order to achieve significant reductions in fuel prices and emissions [2]. In 2019, families consumed around 35% of all electricity in the UK and emitted around 9% of total carbon emissions [3], [4]. The power industry has shifted its focus to renewable energy sources in recent years to lower its carbon footprint during energy generation [5], [6]. Energy policy incentives have aided the installation of distributed energy resources at various levels of the energy system all around the world. The energy policy incentive programs have supported the production of energy at a small level. Some of the European countries have utilized the feed-in tariff (FiT) scheme to store the power that could optimally be used for self-requirement [7]. Due to the deployment of incentive-based programs such as the FiT policies, there has been a considerable growth in the installation of photovoltaic (PV) systems since 2010 [8]. Recent changes in FiT regulations, such as the closure of the Renewable Obligation scheme for small-scale solar PV with a capacity of less than or equal to 5 MW in the UK, will have a significant impact on the scale of household PV installations [9].

According to [10], the cost of battery packs is decreasing, with a 25% reduction in lithium-ion battery packs between 2009 and 2014. The benefits to distribution network operators of maximizing the usage of battery storage for grid-connected home solar PV applications have been studied in [11]-[13]. The influence of changing PV output is minimized by optimizing the operation of battery storage coupled to a residential PV system. The use of a PV power network as a backup supply without storage devices is not a technically viable solution since fluctuating output power impacts grid stability. It's because of substantial variations in weather circumstances, which raise the PV output power's uncertainty. As a result, independent power producers and managing firms, as well as grid balancing authorities, need to anticipate PV output accurately over a wide range of forecast horizons. The precise PV projection will aid power companies in better energy planning and management.

Furthermore, precise forecasting will be advantageous in terms of smart integration of PV generation with the current grid, resulting in increased system reliability [14]. For a 1 MW PV plant in California, authors studied the prediction of PV output performance of five distinct models, namely, k-nearest-neighbor, persistent model, the autoregressive integrated moving average (ARIMA), artificial neural networks (ANNs), and the hybrid genetic algorithm-based ANN model [15]. Using the Fourier series as a predictor, the cyclic behavior of solar radiation may be predicted with a high degree of accuracy, and by combining the different relevant frequencies, this approach may predict sun irradiance. The daily time series profile of solar irradiance can be efficiently built by capturing the annual and intraday cycles, according to [16]. Artificial intelligence-based methodologies were used in the literature to analyze the performance of the trend model. To increase the model's accuracy over a wide range of forecast horizons, certain hybrid models are also used to anticipate solar output load. The authors in [17], developed a hybrid model based on the weather categorization approach and the support vector machine. Clear sky, overcast day, foggy day, and rainy day are the four different types of days. A support vector machine-based forecasting model was created to predict the load forecast for the next 24 hours.

Efficient management of battery storage is critical for many applications, ranging from smart sensors for rechargeable wireless sensor networks [18], [19] to energy storage for power grid [11]. Typically, a single battery storage is used as secondary energy source and occasionally multiple battery storage is operated to achieve application-specific goals. A dual battery storage system was proposed for the efficient utilization of the PV and storage system but suffers from the drawback of the high installation cost of two batteries. The high installation cost of two batteries have become the bottleneck when economic parameters are considered, and there was no information on how much electricity was utilized during the on-peak and off-peak hours. The statistical parameters Skewness and Kurtosis were investigated in order to establish the probability distribution for power import, export, and storage, and this technique has its pros and cons [20]. Authors in [21] presented the unreliability of skewness-based tests; that is, their inability to discriminate between skewed and non-skewed distribution. Estimates of product moments of skewness and kurtosis in a random sample are not robust in the presence of extreme values, and they are also constrained by the sample size. For small samples, these estimates are known to be extremely skewed and have a significant variance, according to [22]. If the value of kurtosis or skewness is too large or too small, there is concern about the distribution's normalcy, [23]. The problem with the shape statistics of skewness and kurtosis is that until thousands of data are involved in the computation, the shape statistics will have so much uncertainty that they will not provide any useful information about which probability models might be reasonable candidates for a process.

The main goal of this study is to propose an optimal scheduling to lessen the cost of electricity purchased and propose a forecasting model to overcome the drawbacks of the Skewness and Kurtosis probability distribution method. One of the main aims of this study is to determine the cost-effectiveness of installing stationary battery storage under time-varying power rates. For this purpose, a set of real half-hourly PV output data and home demand data collected for a year are used to run the scheduling model for the PV system, storage system, and load. A robust, intelligent, and adaptable forecast model is proposed that can account for the factors affecting PV production to improve forecast accuracy. Therefore, LSTM network is utilized for predicting the one-day-ahead consumer load and PV power generation in this study. For three different cases, the cost of electricity during the on peak-hours, off peak-hours, amount of electricity sold, and the net cost of electricity are presented.

The following is a breakdown of the paper's structure. Section II covers the data description. Section III depicts the suggested scheduling, and forecasting by the LSTM network, which is built by taking PV output and load demand into account. Section IV describes the scheduling and forecasting results. Finally, section V brings this study to the conclusion.

## II. Data Description

Half-hourly household power load profiles with a minimum load of 0.213 kW and a maximum load of 0.95 kW were collected from ELEXON for a complete year [24]. The Sheffield solar microgeneration database provided the PV monitoring data [25]. The Sheffield microgeneration database collects data from voluntary PV owners to keep track of PV generation in the United Kingdom (UK). The microgeneration database was randomly picked for half-hour generation from 50 PV systems. The generation data is collected as a series of half-hourly meter readings. The economy 7 tariff is a time-of-use tariff as illustrated in Fig. 1 and offers reduced electricity rates during off-peak hours [26]. The remainder of the day is subjected to a high tariff rate. This tariff is used by around 9% of UK residential power users.

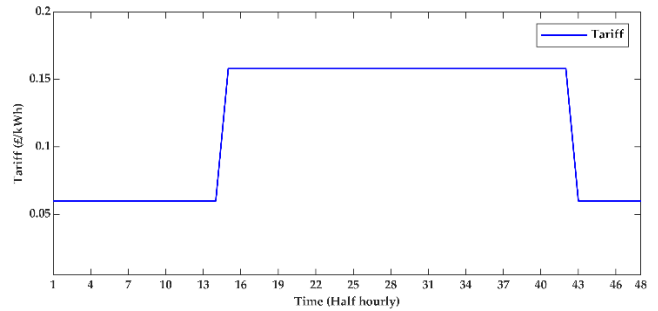

Fig. 1. Economy 7 tariff [26].

## III. Scheduling, and Forecasting by LSTM Network

The contemporary world is experiencing an energy shortage, and this scarcity has prompted academics to experiment with various energy sources. Solar energy is seen as a low-cost and environmentally responsible way to address the energy dilemma. Solar power, on the other hand, has its own set of limitations.

generation is lesser than that of the demand, while the plots in the lower middle and right indicate that the solar PV generation is higher than the demand. Therefore, an optimal scheme is required to utilize solar energy fully to meet the demand. In the case, where the excess is more significant than zero, this power is used to charge the battery for future use, and in the case, where unmet demand is less than zero, it would be advantageous to utilize the power from the storage system. If the storage system is unable to meet demand, power is imported from the grid. Fig. 3 depicts a schematic of our proposed paradigm.

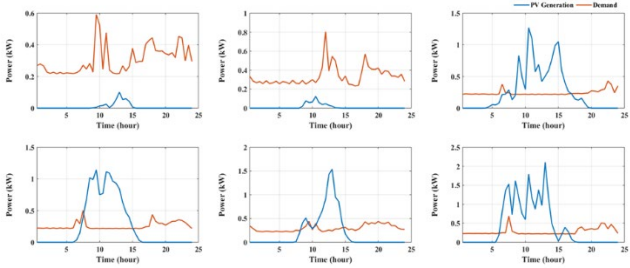

Fig. 2. Power profile for various days in a year.

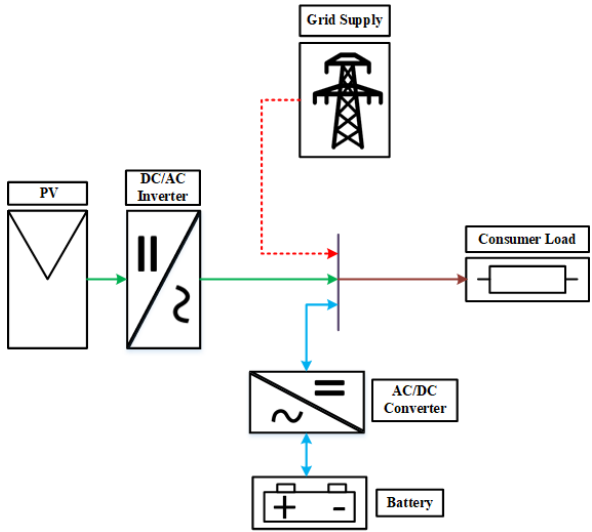

The cost function $J_1$ defined to calculate the electricity charges is shown below:

Subjected to:

$$P_E = \begin{cases} P_S - P_L & if\ P_S > P_L \\ 0 & if\ P_S \leq P_L \end{cases}$$

$$p_t = \begin{cases} p_{i_{tar}} & if\ P_L > P_S \\ p_{e_{tar}} & if\ P_S > P_L \end{cases}$$

where the objective function is indexed by the set *(d, t)*, in which *d* is the set in the domain of $1 \leq d \leq$ 365, *and t is* the set over $0 \leq t \leq 48$, $P_E$ is the difference of the power generated by the PV system $P_S$ and the demand of the consumer $P_L$, $p_{i_{tar}}$ is the purchasing tariff at which the power is bought to meet the unmet demand, $p_{e_{tar}}$ is the selling tariff at which the excess power $P_E$ is sold to the grid.

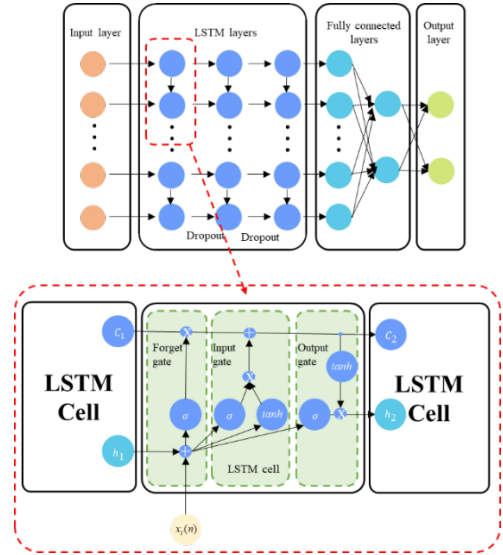

Fig. 4. The architecture of the LSTM network [29].

To achieve higher penetration of solar power technology, it is essential to forecast PV output and it will also help to reduce the reliance on fossil fuels as a source of energy. Furthermore, with precise PV forecast, grid control, power scheduling, unit commitment, and energy management systems may be created effectively. PV output forecasts are classified into four groups based on their forecast horizon: extremely short-term, short-term, medium-term, and long-term PV production forecasts. The accuracy of the PV output forecast is determined by the prediction model's performance. This is due to the individual forecast model's capacity to manage meteorological uncertainties. To reliably anticipate PV output, a number of statistical and AI-based forecast models have been developed. A recursive neural network (RNN) is a type of feed-forward neural network that can be modeled to handle time-sensitive sequences. The RNN takes input from the current state as well as the previous state's hidden layer at any given time. For that timestamp, the output is calculated for the supplied hidden state and is the concealed state memory of RNN. It stores information about earlier data that has been seen by the network. The output at each time is indirectly related to all the preceding inputs due to the relationship between consecutive states. The simple RNN architecture suffers from a vanishing gradient problem in long-range sequences, causing the RNN to forget key information along the chain. By re-parameterizing the RNN, the long short-term memory (LSTM) network solves this problem [27], [28]. The LSTM network's fundamental principle is to manage data flow in the RNN by inserting three gates (forget gate, input gate, and output gate). The input layer, LSTM layers, fully linked layers, and output layer make up the LSTM network. The architecture of LSTM layers is made up of interconnected LSTM cells, each of which is made up of three gates as shown in Fig. 4 [29]. The forget gate adaptively forgets or resets the cell's memory by scaling down the internal state of the cell before adding it as input to the cell through the cell's self-recurrent link. The input gate regulates the flow of input activations into the LSTM cell, while the output gate regulates the flow of cell activations out of the cell to neighboring and related cells. Consequently, the LSTM network's topology keeps the long-term dependency gradient from vanishing.

During supervised learning of the LSTM network, the desired output datum of the LSTM network for predictive power management is of two types, viz, PV power production and consumer load in the following time interval.

## IV. RESULTS AND DISCUSSION

The progress in science and technology has steered us towards the non-replenishing and renewable energy sources to tackle the energy crisis, with low emission of $CO_2$. The utilization of a storage system with PV system serves as a substitute to reduce greenhouse gasses. Fig. 3 shows the schematic diagram of the proposed hybrid PV and the storage system. PV power is prioritized by utilizing the power onsite to meet the demand. When the PV system generates an excess amount, it is stored in the storage system, until the battery reaches a maximum state of charge. The power is sold to the grid at the set rate when the storage system reaches its maximum capacity. Demand is met by the storage system when demand exceeds PV generation or during peak hours. When the battery voltage falls below a certain level, power is purchased from the grid at the specified pace. Fig. 5 depicts the consumer's demand profile for a year.

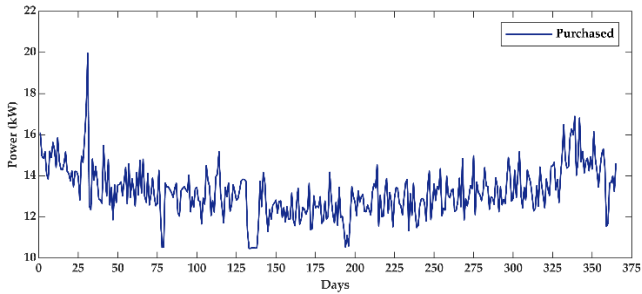

Fig. 5. Power profile without PV and battery system.

When there is no PV or battery storage installed to support the consumer demand during on-peak or off-peak hours, the utility grid is the sole source of supply, and the Economy 7 tariff is in effect. During on-peak or off-peak hours, the power required to meet demand is purchased at the regulated import rate. The highest demand of 19.8 kW is recorded in January, as shown in Fig. 5, which illustrates the grid-purchased electricity for the entire year.

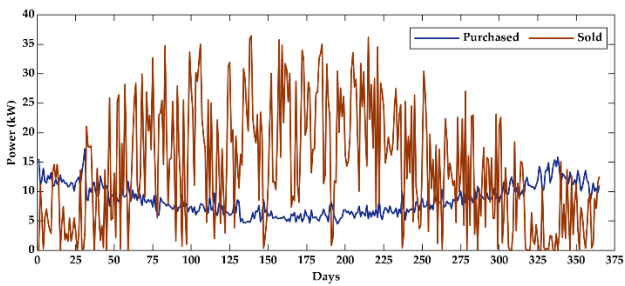

Fig. 6. Power profile with PV and utility grid.

PV–BSS connected to a low-voltage grid is analyzed in this scenario. The need is met by the PV system. If there is excess power, it is stored in the battery or exported to the grid at the 0.04 £/kWh export rate. If there is an unmet need, the battery will deliver electricity during peak hours only.

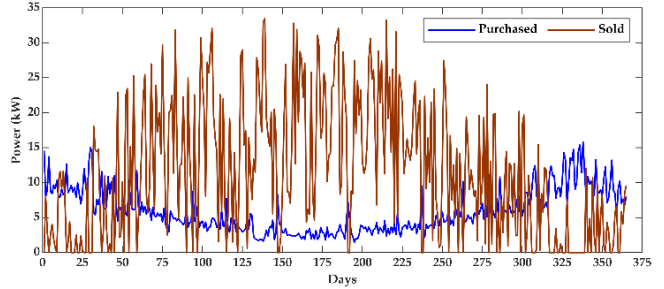

Fig. 7. Power profile with PV, utility grid, and battery.

Table I shows the cost analysis of the various cases with the proposed methodology.

TABLE I. COST ANALYSIS OF DIFFERENT CASES

| Cases | Import cost (£) | | Export cost (£) | Net cost (£) |
|---|---|---|---|---|
| | Peak hour cost (£) | Off-peak hour cost (£) | | |
| 1) Utility grid only | 259.96 | 122.84 | 0 | 384.24 |
| 2) PV and utility grid | 111.68 | 52.535 | 102.5 | 61.69 |
| 3) PV, battery, and utility grid | 53.12 | 35.413 | 50.41 | 38.14 |

The LSTM network's training parameters are as follows: The ADAM optimization algorithm is employed with a learning rate of 0.001, and a total number of training epochs of 50 and batch size is 1. Loss function is the MSE. Inadequate learning rates may result in a local minimum and overfitting, both of which are undesirable. To avoid local minima, dropout and gradient are utilized. LSTM network's training is performed by randomly splitting the whole dataset into training of 80% dataset and validation of 20 % dataset.

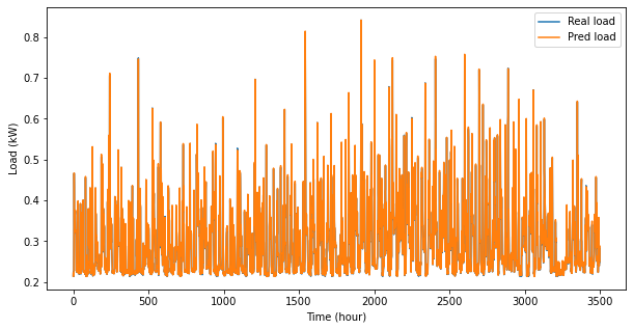

Fig. 8. Prediction of one-day load using LSTM network.

LSTM network's input are 29 prior days and output is the 1 next day for PV power and load. In Fig. 8 and Fig. 9, the forecasted consumer load and PV power follow the same trajectory of real consumer load and PV data with the MSE of 0.000012, root mean square error (RMSE) of 0.003560, and $R^2$ of 0.999379. The results of the proposed model show that, the reduction of electricity cost is 37.5 % when PV and utility grid are utilized, and is 43.06 % when PV, utility grid, and storage system are utilized.

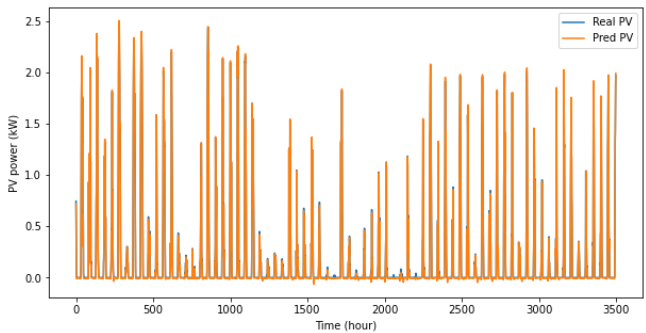

Fig. 9. Prediction of one-day PV power using LSTM network.

## V. Conclusion

Solar energy is considered as a green and everlasting energy source in the power sector. The advancement of renewable energy technology has induced certain obstacles, and the best way to estimate the accurate output supplied by these technologies is to introduce new machine learning approaches. In this study, we proposed the scheduling and forecasting of hybrid PV and storage system by LSTM network. The proposed system has been operated optimally, by utilizing the stored power during peak hours. The analysis shows that utilizing the proposed peak-hour strategy resulted in the reduced electricity bill and provides incentives and attraction to users for the installation of hybrid systems benefitting from the FiT scheme. The LSTM network used improves prediction accuracy while reducing computational time and mistakes. The LSTM network regressor's performance results showed the mean MSE of 0.000012, RMSE of 0.003560, $R^2$ of 0.999379, training time of 5,345 seconds, and a calculation time of 0.02 seconds. The reduction of electricity cost was 37.5 % when only PV and utility grid were utilized, and it was 43.06% when PV, utility grid, and storage system are utilized.

## Acknowledgment

This work was supported by the Korea Institute of Energy Technology Evaluation and Planning (KETEP) and the Ministry of Trade, Industry & Energy (MOTIE) of the Republic of Korea (No. 20192010107290).